\magnification 1200
\hsize=31pc             
\vsize=55 truepc         
\baselineskip=26 truept 
\hfuzz=2pt               
\vfuzz=4pt               
\pretolerance=5000        
\tolerance=5000        
\parskip=0pt plus 1pt 
\parindent=16pt       
\def\avd#1{\overline{#1}}
\def\avr#1{\overline{#1}}
\def\avt#1{\langle#1\rangle}

\def\s{\sigma}
\def\S{{\bf s}}

\def\b { \beta}

\vskip 1.truecm\noindent
\centerline {\bf Beyond the  mean field approximation for spin glasses}
\vskip 1.4truecm
\noindent
\centerline{ Maurizio Serva$^{1}$ and  Giovanni Paladin$^{2}$}
\vskip .4truecm
\centerline{\it $^{1}$
 Dipartimento di Matematica,  Universit\`a dell'Aquila and  I.N.F.M.}
\centerline{\it I-67100 Coppito, L'Aquila, Italy}
\bigskip
\centerline{\it $^{2}$
 Dipartimento di Fisica,  Universit\`a dell'Aquila and I.N.F.M.}
\centerline{\it I-67100 Coppito, L'Aquila, Italy}
\vskip .7truecm
\centerline{ABSTRACT}
\vskip .4truecm
We study the $d$-dimensional random Ising model 
using a suitable type of Bethe-Peierls approximation 
 in the framework of the replica method. 
We take into account the correct interaction only 
 inside replicated clusters of spins.
 Our ansatz is that the  interaction of the borders of the clusters
with the external world can be described via 
 an effective interaction among replicas.
The Bethe-Peierls model can be mapped
into  a single Ising model with a random gaussian field,
 whose strength (related to the effective coupling between two
 replicas) is determined via a self-consistency equation.
  This allows us to obtain
analytic estimates of the internal energy 
 and of the critical temperature in $d$ dimensions. 
\noindent
\medskip
PACS NUMBERS: 05.50.+q, 02.50.+s 
\vfill\eject

\bigskip
\bigskip

{\bf Introduction}
\medskip
The mean field solution and its 
improvements, such as the Bethe-Peierls approximation [1,2], 
give good approximations of the
 critical temperature and of the internal energy 
 in  many statistical models. 
 We show that the same methods can be applied 
to spin glasses by considering the
 overlap among replicas instead of the magnetization.
 Indeed, the  appropriate ansatz for spin glasses
 is assuming that the effect of the thermal bath 
 on a replicated cluster of neighbors  spins
  produces an effective coupling among replicas.
 We shall give an {\it a posteriori} justification of such a hypothesis
by proving that the Sherrington-Kirkpatrick (SK)
 model [3] is recovered in the limit of infinite dimension.

This paper  considers the Ising model
 with independent random nearest neighbor coupling 
$J_{ij}$ in absence of external magnetic field. 
 Our main result is that, in  the 
 Bethe-Peierls
 approximation, this model is equivalent to
 a single Ising model with a  random gaussian field whose strength is 
 related to the effective coupling between two replicas.
We thus obtain an estimate of the  internal energy 
 and of the critical temperature in any dimension.

In section 1,  we introduce the Bethe-Peierls ansatz 
 for spin glasses in the framework of the replica method.

In section 2, we prove that this ansatz leads to
 the SK model when $d \to \infty$

In section 3, we show that,  under the hypothesis of no replica symmetry
 breaking,   the $d$-dimensional model can be mapped 
into a single  Ising model with 
random gaussian field. 
We also explicitly compute the replica symmetry solution
 for the internal energy  in $d$ dimensions
 for the spin glass with dichotomic random coupling
 $J=\pm1$. 

In section 4, we show that our method allows to compute
 in a simple way the critical temperature $T_c(d)$.
 The result is very accurate at high dimension.

In section 5, we discuss the possibility to
 use our ideas to implement a clever numerical scheme 
 for determining internal energy and critical temperature
 of $d$-dimensional spin glasses. 

\bigskip
\noindent
{\bf 1. Bethe-Peierls ansatz for spin glasses}
\medskip
\noindent
The partition function 
 of the  Ising models 
 on a  lattice of $N$ sites
 with nearest neighbor couplings $J_{ij}$ which are
independent identically distributed random variables,
 in absence of external magnetic field, is 
$$
Z_N(\beta, \{J_{ij}\})  =  \sum_{ \{ s \} }  \prod_{(i,j)} 
\exp (\beta J_{ij} \sigma_i \sigma_j )
\eqno(1.1)
$$
where the sum runs over the $2^N$ spin configurations $\{s\}$, and 
 the product over  the $d \, N$  nearest neighbor sites $(i,j)$.

In the thermodynamic limit
 almost all disorder realizations has
 the same free energy, i.e the quenched free energy
$$
f=- \lim_{N \to \infty} {1\over \beta N} \avd{\ln Z_N}
\eqno(1.2)
$$
where $\avd{A}$ indicates the average of an observable $A$
 over the distribution of the random coupling $P(J_{i,j})$ .
 In the following we assume that the $P(J_{ij})$ 
 is such that $\avd{J_{ij}}=0$ and  $\avd{J_{ij}^2}=1$.

  On the other hand, it is trivial to compute
the so-called annealed free energy
$$
f_a=-\lim_{N \to \infty} {1\over \beta N} \ln \avd{ Z} \ , 
\eqno(1.3)
$$
corresponding to the free energy of a system where the random coupling
 are not quenched but can thermalize with a relaxation time comparable
 to that one of the spin variables. 
For instance, in  the case of 
 dichotomic random coupling $J_{ij}=\pm 1$ with equal probability, one has 
$$
f_a= - \beta^{-1} \,  (\ln 2 +d \ln  \cosh \beta  \, )
\eqno(1.4a)
$$
while for gaussian coupling, i.e. $P(J_{ij})=
 e^{J_{ij}^2/2} \, / \, \sqrt{2\pi}  \,$, one has
$$
f_a= - \beta^{-1} \, (\, \ln 2 \,  + \,  {\beta^2 d \over 2} \, )
\eqno(1.4b)
$$
However, $f_a$ is in general very different from the quenched free energy.
In order to compute (1.1), it is convenient to use the replica trick [4]. 
Let us thus consider $n$ non-interacting replicas of the disordered 
 system labelled by $\alpha=1,\cdots,n$. 
The corresponding partition function is 
$$
 Z^n=
       \sum_{\{\S\}} 
\exp \left( {\b \sum_{\alpha=1}^n \sum_{(i,j)} J_{ij}
 \s_i^{(\alpha)} \, \s_j^{(\alpha)}} \right)
\eqno(1.5)
$$ 
where the sum runs over the $2^{Nn}$ spin configurations $\{\S\}$
of the replicas, 
$$
\{\S\}   \equiv \{s^{(1)}\},\cdots,\{s^{(n)}\}
$$
with
$$
s^{(\alpha)}= ( \sigma_1 ^{(\alpha)} , \sigma_2^{(\alpha)},
\dots , \sigma_N^{(\alpha)} )
$$

After having performed the average 
$\avd{Z^n}$  and 
found  an analytic continuation at real $n$-values, 
the quenched free energy is given  by 
$$
\avd{\ln \, Z }=\lim_{n \to 0} {1 \over n}\,  \ln \, \avd{Z^n} 
\eqno(1.6)
$$
Even in two dimensions, there is no exact solution
 for this problem. The first non-trivial approximations
 of the quenched free energy can be obtained 
 either by constrained annealed average [5]  or  by
improved mean field approximations of the Bethe-Peierls type.
 Recently we have introduced such an approximation 
 in the dual lattice made of square plaquettes in two dimensions 
 [6]. 
 However, there is no solution of the self-consistency equation 
 at low temperature and it is not trivial to generalize the approach at 
higher dimensions.
 For systems with diluted quenched disorder,
 a different type of improved Bethe-Peierls approximation
 (the Cluster variation method) has been studied in [7]
 without using the replica approach.

 In this paper, we want to work directly on the real lattice,
 by taking into account the correct interactions 
 inside  a pile of replicated cluster made of 
a central spin $\s_0$ and of its $2d$ nearest neighbor $\{\s_k\}$,
 and by considering  only an effective interaction with the external
 world. 
Note that  in the $2d$ case, the clusters are crosses made of $5$ spins.

Separating the two contribution (crosses plus external world)
 in the partition function, we get 
$$
\avd{ Z^n}=
       \sum_{\{\S_{cr}\}} \, 
 \left( \, 
 \avd{ 
\exp( \, 
 \b \sum_{\alpha=1}^n \sum_{k=1}^{2d} \,  J_{k}
 \s_0^{(\alpha)} \, \s_k^{(\alpha)}
)
} \, 
      \sum_{\{\S_{ext}\}} 
\avd{
\exp ( \,  \b \sum_{\alpha=1}^n \sum_{(ij) \neq (0 k)} J_{ij}
 \s_i^{(\alpha)} \, \s_j^{(\alpha)} ) 
} \, 
\right)
\eqno(1.7)
$$ 
where $$J_{k}\equiv J_{0k}$$
are the coupling between the central spin of the cross and 
 its neighbors on the border. 
  The first sum in (1.7) runs over the $2^{(2d+1)n}$ 
spin configurations $\{\S_{cr}\}$
of the replicated crosses labelled by 
($\s_0^{\alpha}, \s_1^{\alpha},\cdots, \s_{2d}^{\alpha}$ )
with $\alpha =1, \dots ,n$
 while the second sum over all the other spins.
The expression obtained by computing the second sum
depends only on the $2dn$ lateral spins $\s_k$. 
The correct Bethe-Peierls ansatz is given by the assumption 
that the interaction among the lateral spins of 
 the replicated crosses and the external word
forces an effective interaction among different replicas
 with a constant $\mu_{\alpha  \beta}$
 that should be determined via a self-consistency equation.
In other terms,  our Bethe-Peierls ansatz is
$$
      \sum_{\{\S_{ext}\}} 
\avd{
\exp \left( {\b \sum_{\alpha=1}^n \sum_{(ij \neq 0k)} J_{ij}
 \s_i^{(\alpha)} \, \s_j^{(\alpha)}} \right)
}
=     K(\beta)  \exp \left( { 
\sum_{\alpha>\beta} \mu_{\alpha \beta}\sum_{i=k}^{2d}
  \s_k^{(\alpha)} \, \s_k^{(\beta)}
} \right)
\eqno(1.8)
$$ 
where $K(\beta)$ is a multiplicative constant which depends on the
temperature but not on the lateral spins.
One expects that 
$\mu_{\alpha  \beta}=0$  in the high temperature phase,
 while it must have a non-zero value
 in the glassy phase. 

Therefore, instead of the (1.7), 
 we have to compute an effective  partition function $Z_n$
$$
Z_n= \sum_{\{\S_{cr}\} }  \prod_{k=1}^{2d} \, 
\avd{
\exp 
\left( 
\b \sum_{\alpha=1}^n  J_{k}
\s_0^{(\alpha)} \, \s_k^{(\alpha)} 
\right)
} 
\, 
\exp 
\left(
\sum_{\alpha>\beta} \mu_{\alpha \beta}
\s_k^{(\alpha)} \, \s_k^{(\beta)} 
\right) 
\eqno(1.9)
$$ 

A further simplification can be reached for 
 dichotomic coupling $J_{ij}=\pm1$ where one can perform
 the gauge transformation $\s_k^{(\alpha)} \ \to J_{k} \,  
\s_k^{(\alpha)} $ on the lateral spins, leaving 
 the free energy unchanged. In this case
the averaged partition function (1.9) becomes 
$$
Z_n =
       \sum_{\{\S_{cr}\} }  \prod_{k=1}^{2d} \, 
\exp \left( \b \sum_{\alpha=1}^n  
 \s_0^{(\alpha)} \, \s_k^{(\alpha)} \right)  \, 
\exp \left( 
\sum_{\alpha>\beta} \mu_{\alpha \beta}
  \s_k^{(\alpha)} \, \s_k^{(\beta)} \right) 
\eqno(1.10)
$$ 
This relation  implies the rather surprising 
 result that a non-disordered Ising system exhibits the same behavior
 of a spin glass { \it if} one imposes 
the appropriate interaction among different replicas.

At this point, the effective coupling $\mu_{\alpha  \beta}^*(\beta)$
   is given by the self-consistency equation
$$
 \lim_{n \to 0 } \avt{  \s_k^{(\alpha)} \, \s_k^{(\beta)}}_n =
\lim_{n \to 0}   \avt{  \s_o^{(\alpha)} \, \s_0^{(\beta)}}_n 
\eqno(1.11)
$$
where $\avt{\cdot}_n$ represents the thermal average over the replicated 
system. 
Then, the Bethe-Peierls estimate of the 
  internal energy is 
$$
U_{BP}(\beta)= \lim_{n \to 0} 
-{1 \over 2n} \left[
 {\partial \over \partial \beta} \,  \ln 
Z_n
(\mu_{\alpha  \beta}, \beta) \right]_{\mu_{\alpha  \beta}=\mu_{\alpha \, 
\beta}^*}
\eqno(1.12)
$$
Let us 
anticipate that the Bethe-Peierls approximation predicts 
 a phase transition at a critical temperature $T_c(d)$
above which  $\mu^*_{\alpha \b}=0$.
As a consequence the Bethe-Peierls solution
 coincides with the annealed one in the high temperature phase, i.e.
$$
U_{BP} = {d\over d\b} \, \left[ \b f_a(\b) \right] 
\qquad{\rm for} \ 
\b < \b_c
\eqno(1.13)
$$
\bigskip
\noindent
{\bf 2. The infinite dimensional limit}
\medskip
\noindent
The model defined by (1.9) becomes the infinite range
 SK model [3] in the limit $d \to \infty$. This result has a great
 importance since provides a good evidence that we have chosen the correct
 Bethe-Peierls ansatz for spin glasses. 
 In this section, we prove that the self-consistency equation (1.11) 
 in the limit $d \to \infty$ gives the equation for the overlap
 of the SK model.

Let us recall that the averaged partition function of 
 the infinite range SK model after some simple algebraic manipulation
 becomes 
$$
\avd{Z^n}=(\avd{Z})^n  \ \left[  \max_{q_{\alpha  \beta}} \, 
   {1 \over 2^n}       \sum_{ \{ \sigma \} } 
\exp \b^2  \left(  \sum_{\alpha>\beta}
 q_{\alpha  \beta} \s^{(\alpha)} \, \s^{(\beta)}
-{q_{\alpha  \beta}^2\over 2} \right)\right]^N
\eqno(2.1)
$$
where the sum is on the $2^n$ realizations $\{\sigma \}$
of the $n$ spins $\sigma^{(1)},\dots ,\sigma^{(n)}$. 
In the high temperature phase $T \ge T_c$, one has $ q_{\alpha  \beta}=0$
so that $\avd{Z^n}=(\avd{Z})^n$, while in the glassy phase
one has a non trivial overlap $q_{\alpha  \beta}=q_{\alpha  \beta}^*(T)$
which maximizes $(\avd{Z})^n$, that is
$$
q_{\alpha  \beta}=\avt{ \s^{\alpha} \s^{\b} }  \equiv
 {
 \sum_{ \{ \sigma \} } \s^{(\alpha)} \, \s^{(\b)} \, 
\exp \left( \b^2 \sum_{\alpha>\beta}
 q_{\alpha  \beta} \s^{(\alpha)} \, \s^{(\beta)}
\right)
 \over
\sum_{\{\sigma\}} 
\exp \left( \b^2 \sum_{\alpha>\beta}
 q_{\alpha  \beta} \s^{(\alpha)} \, \s^{(\beta)}
\right)
}
\eqno (2.2)
$$

 In order to get the correct $d \to \infty$ limit of 
the self-consistency equations (1.11), 
we should use the rescaling
 $$
\beta \to {\b \over \sqrt{2d}} \qquad \qquad
\mu_{\alpha  \beta} \to \b^2 \mu_{\alpha  \beta}
\eqno(2.3)
$$
Now, the disorder average in $Z_n$ 
is easily performed since at large $d$ the first exponential
 in (1.9) can be expanded  in Taylor series up to the second order, so that
$$
 \avd{  \exp(\b \,  (2d)^{-1/2}  J_k S_k }
=
 \avd{ 1+\b (2d)^{-1/2} J_k  S_k \, 
+ {\b^2 \over 4d}\,  J_k^2 \, S_k^2 }
+O(d^{-3/2})=
$$
$$
=
  1+{\b^2 \over 4d} S_k^2+O(d^{-3/2})
 =\exp({\b^2\over 4d} S_k^2)+O(d^{-3/2})
$$
where $S_k\equiv \sum_{\alpha}  \s_0^{(\alpha)} \, \s_k^{(\alpha)}$. 
 The distribution of the coupling is irrelevant provided that
$\avd{J}=0$ and $\avd{J^2}=1$.
 Therefore, 
a part small correction $O(d^{-3/2})$, the partition function (1.9)
becomes 
 $$
Z_n=
\sum_{ \{ \S_{cr} \} }  
\exp \b^2 \, \left( \, \sum_{\alpha>\beta} 
 \  \s_0^{(\alpha)} \, \s_0^{(\b)} \
 \, {1\over 2d} \sum_{k=1}^{2d} \s_k^{(\alpha)} \, \s_k^{(\b)} 
 \, 
 +  \mu_{\alpha  \beta} \sum_{k=1}^{2d} \s_k^{(\alpha)} \, \s_k^{(\beta)}
\right)
\eqno
(2.4)
$$ 
implying that
$$
<\s_k^{(\alpha)} \, \s_k^{(\b)}>=
{
\sum_{ \{ \sigma_k \} }   \s_k^{(\alpha)} \, \s_k^{(\beta)} \, 
\exp \left(
 \b^2 \sum_{\alpha>\beta}
 \mu_{\alpha  \beta} \s_k^{(\alpha)} \, \s_k^{(\beta)}
\right)
\over 
\sum_{ \{ \sigma_{k} \} }   
\exp \left(
 \b^2 \sum_{\alpha>\beta}
 \mu_{\alpha  \beta} \s_k^{(\alpha)} \, \s_k^{(\beta)}
\right) 
}\ + \ O(d^{-1/2})
\eqno(2.5)
$$ 
where $k$ is one of the lateral sites  of the 
$d$-dimensional crosses and the sum is on the $2^n$ 
realizations ${\sigma_k }$ of the $n$ spins 
$\sigma_k^{(1)},\dots ,\sigma_k^{(n)}$. 

On the other hand, in the limit $d \to \infty$,
 the corresponding relation for the central spins
 can be written as 
$$
\avt{\s_0^{(\alpha)} \, \s_0^{(\b)}} \, 
= {
\sum_{ \{ \sigma_{0} \} }   
\s_0^{(\alpha)} \, \s_0^{(\beta)} \, 
\exp( \,
 \b^2 \sum_{\alpha>\beta} \, 
\avt{  \s_k^{(\alpha)} \, \s_k^{(\beta)} } \, 
 \s_0^{(\alpha)} \, \s_0^{(\beta)}
\, )
\over 
\sum_{ \{ \sigma_{0} \} }   
\exp (\,  \b^2 \sum_{\alpha>\beta}
\avt{  \s_k^{(\alpha)} \, \s_k^{(\beta)} } \, 
 \s_0^{(\alpha)} \, \s_0^{(\beta)}
 \, )
}
\eqno(2.6)
$$
since one has 
$$
<\s_k^{(\alpha)} \, \s_k^{(\b)}>=
\lim_{d \to \infty} {1\over 2d} \, 
\sum_{k=1}^{2d} \s_k^{(\alpha)} \, \s_k^{(\beta)} 
\eqno(2.7)
$$
A direct comparison of (2.6) and (2.5), shows that the self-consistency 
equation (1.11) is satisfied only if
$$
\mu_{\alpha  \beta}=<\s_k^{(\alpha)} \s_k^{(\b)}> 
$$
that is the equation for the overlap of the SK model.
 We can thus identify 
 the coupling $\mu_{\alpha \beta}$ with the overlap 
 $q_{\alpha \beta}$ for $d \to \infty$.
\bigskip
\noindent
{\bf 3. Replica symmetry solution in the Bethe-Peierls approximation}
\medskip
\noindent
It is possible to obtain the
replica symmetry solution of a $d$-dimensional spin glass
 in the Bethe-Peierls approximation.
We must note that in the case $\mu_{\alpha \b}=\mu$,
 the averaged partition function (1.9)
of $n$ replicated crosses is
$$
Z_n =
       \sum_{\{\S_{cr}\} }  \prod_{k=1}^{2d} \, 
\avd{
\exp \left( \b J_{k} \, \sum_{\alpha=1}^n  
 \s_0^{(\alpha)} \, \s_k^{(\alpha)} \right)
} 
 \ \exp \left( \mu
\sum_{\alpha>\beta}
  \s_k^{(\alpha)} \, \s_k^{(\beta)} \right)
\eqno(3.1)
$$ 
A part constant multiplicative factors, it can also be written as
$$
Z_n =
       \sum_{\{\S_{cr}\} }  \prod_{k=1}^{2d} \, 
\avd{
\exp \left( \b J_{k} \, \sum_{\alpha=1}^n  
 \s_0^{(\alpha)} \, \s_k^{(\alpha)} \right)
} 
 \ \exp { \mu  \over 2} \left(   
\sum_{\alpha}
  \s_k^{(\alpha)} \right)^2
\eqno (3.2)
$$ 
that is  bilinear in $\s_k$ .
 In order to linearize (3.2), we should use 
the standard gaussian identity
$$
\exp(x^2/2)=
{1\over \sqrt{2\pi} } 
\int_{\infty}^{\infty} d\omega \, \exp(-\omega^2/2) \, \exp(\omega x)
$$
so that one has 
$$
Z_n  =       \sum_{\{\S_{cr}\} }  \prod_{k=1}^{2d} \, 
\avr{ \,
\exp \left( \, \sum_{\alpha=1}^n   ( \,  \b \, J_k \, 
 \s_0^{(\alpha)} \, \s_k^{(\alpha)} 
+ \, \sqrt{\mu}    \, \omega_k \,   \s_k^{(\alpha)} \, ) \, 
\right) \,  } 
\eqno(3.3)
$$ 
where
 $$
\avr{\psi} = {1\over (2\pi)^{d/2}} 
 \int \cdots \int \prod_{k=1}^{2d}
d\omega_k 
e^{-{\omega_k^2/2} } \prod_{k=1}^{2d}
P(J_k) \, dJ_k  \ \psi
\eqno(3.4)
$$
 indicates now the average over the standard 
 gaussian variables $\omega_k$ and 
over the coupling $J_k$ between central and lateral spins.

This transformation has the advantage to allow for a
 factorization of the product over the replicas in (3.3),
implying that
$$
Z_n=\avr{\Phi^n}
\eqno(3.5)
$$
with
$$
\Phi=
\sum_{\{\S_{cr}\} }  \prod_{k=1}^{2d} \, 
\exp \left( \b \, J_k \, 
 \s_0 \, \s_k \,  
+ \, \sqrt{\mu}   \, \omega_k  \s_k
\right) 
\eqno(3.6)
$$
This is the main result of the section.
It establishes that, in the Bethe-Peierls approximation,
the replica symmetry solution is equivalent
 to that one of  a single Ising model with
 a random gaussian field applied to the boundaries of  the cross. 
 This field has a strength related  to the coupling among replicas and 
 describes the interaction of the cluster of $2d+1$ spins
 with the external world. 
 
The explicit sum over the lateral spins $\s_k$'s gives
$$
Z_n =  
\avr{ \, \left(
        \sum_{\{s_{cr}\} } W_{\mu}(\s_0)  \,  \right)^n \,  } 
\eqno(3.7)
$$ 
where $W_{\mu}$ is the non-normalized weight of the central spin
$$
W_{\mu}(\s_0)=\prod_{k=1}^{2d} \, 
 2 \cosh (\b  J_k \, \s_0 \, 
+ \, \sqrt{\mu}    \, \omega_k)
\eqno(3.8)
$$
obtained after summing over the configurations of
 the $2d$ lateral spins $\s_k$.
The probability of the central spin thus is
$$
P_{\mu}(\s_0)={W_{\mu}(\s_0)\over W_{\mu}(\s_0=1)+  W_{\mu}(\s_0=-1)  }
\eqno(3.8)
$$
and    is itself a random quantity depending on the $2d$ random gaussian 
fields and to the $2d$ random coupling $J_k$.

Because of the replica symmetry, 
the self-consistency equation (1.11) 
 for determining $\mu^*$, and so the needed strength 
of the random field, assumes the simpler form
$$
\avr{<\s_0>^2}=\avr{<\s_1>^2}
\eqno(3.9)
$$
where the thermal 
average of the central spin is 
$$
<\s_0>=
 \sum_{\s_{0}=\pm 1} \s_0 \ P_{\mu}(\s_0)
\eqno(3.10)
$$
and the thermal average of one of the lateral spins is 
$$
<\s_1> =
         \sum_{\s_{0}=\pm 1}   
 \tanh(\b J_k  \s_0 \, + \, \sqrt{\mu}    \, \omega_1)
 \   P_{\mu}(\s_0)
\eqno(3.11)
$$
In order to find the internal energy we have to compute
$$
\lim_{n \to 0} {1 \over n}\ln Z_n =  
\avr{ \, \ln
        \sum_{\s_{0}=\pm 1}  \prod_{k=1}^{2d} \, 
 2 \cosh (\b  \, J_k \, \s_0 \, 
+ \, \sqrt{\mu}    \, \omega_k)
\,  } 
\eqno(3.12)
$$
and then, following (1.12),
 the internal energy is given by a derivative 
 at $\mu=\mu^*$, solution of (3.9)
$$
U_{BP}(\b)=-d \, \avr{
 \sum_{\s_{0}=\pm 1}  J_k \s_0
 \,  \tanh(\b  J_k \s_0 \, + \, \sqrt{\mu}    \, \omega_1) 
 \ P_{\mu}(\s_0)
} 
\eqno(3.13)
$$

It is worth stressing that in the limit $d \to \infty$,
 the self-consistency equation (3.9) becomes,
by virtue of the results of section 2, 
$$
\mu=\lim_{d \to \infty} \avd{<\s_1>^2}
\eqno(3.14)
$$
The above expression, after performing
the rescaling (2.3)   gives
the replica symmetry solution for the overlap of the
 SK model in the glassy phase,
$$
\mu=\avr{ \tanh^2(\b    \, \omega \, \sqrt{\mu} )  } 
\eqno(3.15)
$$
where $\omega$ is again a standard gaussian.

For the $\pm J$ model, 
after the gauge transformation $J_k\s_k \to \s_k$, 
 the probability $P_\mu(\s_0)$ depends
 only on the gaussian fields and is independent of the
 coupling $J_k$.
 We can thus put $J_k=1$ in the formulas from (3.3)
 to (3.13) and the average (3.3) should be taken 
 only over  the $2d$ gaussian variables $\omega_k$'s.

Fig 1 shows the replica symmetry solution $ T^2 \mu^*(T) \, / 2 d \,$ 
 as function of the rescaled temperature
 $ T/\sqrt{2d}$ at $d=2\, , 3\, , 4,6$ for the $\pm J$ model.
 The effective replica coupling 
 $\mu^*$ vanishes above the critical temperature $T_c(d)$ as we shall discuss
 in the next section.
 As a consequence the
 internal energy $U_{BP}$ is equal to the annealed internal energy
 at $T \ge T_c$.

Another peculiar effect arises in the low temperature regime
 where $T^2 \, \mu^*(T) $ decreases below $T_0 \approx 0.5$.   
This vanishing of the coupling  $T^2 \, \mu^*(T) $ 
 can be understood by the following qualitative arguments.
Consider the model defined by (3.6) with $J_k=1$ which is
 originated by the $\pm J$ model.  On the lateral spins,
 there is a competition between the coupling with the random field 
  $\sqrt{\mu} \,   \omega_k \s_k$
 and the ferromagnetic interaction $\beta \s_0 \s_k$.
 It gives origin to a frustration
of the system below  $T_c$.
 However, if the temperature is very low,
the ferromagnetic interaction dominates since the work
 $\sqrt{\mu^*} /\b$ 
necessary to win the tendency of the spin $\s_k$
to align with the field vanishes. In correspondence
 the system would become ferromagnetic with a ground state 
$U_0=-d$ as it happens for the annealed model. 
 Such a regime is clearly unphysical,
and one can trust in our results only 
when  the work  $\sqrt{\mu^*} /\b$  made to destroy
 the long range order in the glassy phase
 is a non-increasing function of the temperature,
 i.e. for $T \ge T_0$.

Notice that the mean field coupling is obtained 
at increasing $d$, by the {\it `a posteriori'} rescaling
 $\mu^* \to \mu^*/(2d\b^2)$
 for the coupling and $T \to T'\equiv T \, / \, \sqrt{2d}$
  for the temperature
in the sequence of lines for the $d$-dimensional functions.
It is therefore natural our choice for the coordinates of 
 Fig 1.  Let us also stress that the rescaled temperature 
 below which $T^2 \, \mu^*$ decreases  is $T_0'=0.5 \, / \, \sqrt{2d}$ so that 
 the unphysical ferromagnetic regime disappears when $d \to \infty$.

The internal energy $U_{BP}(T)/d$ is shown in Fig 2
 for the $\pm J$ model at $d=2\, , 3,\, 4 \, ,6$. 
 An estimate of the ground state  energy $U_0$ 
 can be obtained by $U_{BP}(T_0)$
 as previously argued.
 Using this hypothesis we get
\medskip
\noindent
$U_0 = -1.51$ at $d=2$,  \hfill\break
$U_0 = -1.88$ at $d=3$, \hfill\break
$U_0 = -2.204$ at $d=4$, \hfill\break
$U_0 =-2.718$ at $d=6$\hfill\break
At $d=2$ we can compare our analytic estimate 
 with the numerical result [8], $U_0=-1.404$. 

 It is an open issue to understand whether better 
  estimates can be obtained via (1.12)  
 with a replica symmetry breaking solution $\mu^*_{\alpha \beta}$.
\bigskip
\noindent
{\bf 4. Phase transition and critical temperature in finite dimension}
\medskip
\noindent
 The Bethe-Peierls method and its improvements are 
able to give accurate estimates of 
  the critical temperature in a disordered system.
Using  a replica symmetry approach, good analytic results have been
 obtained for diluted spin glasses [9] and other randomly
 frustrated systems with finite connectivity [10]. 

In the framework of the results of the previous section,
we should note that  
at the transition point,  the order parameter $\mu^*$ vanishes.
Therefore, the critical temperature can be computed
from (3.3) considering only the first order 
of its expansion in $\mu^*$.

Let us first compute the thermal average of
the central spin
$$
<\s_0> \ = \
\sqrt{\mu} \ \tanh (\b \, J_k) \ \sum_{k=1}^{2d} \omega_k
+O(\mu)
\eqno(4.1)
$$ 
Since this expression appears in (3.3) only in a squared form
it is not necessary to compute higher orders than $\sqrt{\mu}$.
Analogously, the thermal average of one of the lateral spins is
$$
<\s_1> \ = \
 \sqrt{\mu} \, \tanh (\b \, J_1) \, \sum_{k=2}^{2d} 
 \tanh^2 (\b \, J_k) \, \omega_k
+ \sqrt{\mu} \ \omega_1 +O(\mu)
\eqno(4.2)
$$
Inserting this expression in the consistency
equation (3.9) one obtains:
$$
\mu \, 2d \, \avd{\tanh^2 (\b J)} = \mu \, (2d-1) \,
\avd{\tanh^2 (\b J)} \left(\avd{\tanh^2 (\b J)} \right)^2  +\mu
\eqno(4.3)
$$
where $J$ is one of the couplings.
Eq. (4.3) gives the critical temperature $T_c=\b_c^{-1}$ 
as a function of the dimension 
$$
\avd{\tanh^2 (\b_c J_k)}  = {1 \over 2d-1}
\eqno(4.4)
$$
In the case of the $\pm J$ model, this equation becomes
$$
\tanh^2 (\b_c)
 = {1 \over 2d-1}
\eqno(4.5)
$$
In Fig 3, we compare the Bethe-Peierls critical temperature (4.5)
with the numerical result obtained in the literature 
 for the $\pm J$ model [11].

We have also computed the critical temperature of the gaussian model
via a numerical solution of (4.4).
 In this case:\medskip
\noindent 
 $T_c=1.19$ at $d=2$,\hfill\break
 $T_c=1.81$ at $d=3$ (numerical result $T_c=1.0$),\hfill\break
 $T_c=2.28$ at $d=4$ (numerical result $T_c=1.8$),\hfill\break
 $T_c=2.67$ at $d=5$,  \hfill\break
 $T_c=3.06$ at $d=6$. 

 Let us remark that $T_c$ is finite in $d=2$.
 This spurious transition is a typical 
and well-known effect of mean field approximations.
 In fact, The Bethe-Peierls approximation, 
 gives a lower critical dimensionality $d_c=1$, where $T_c=0$,
 while there is a good numerical evidence 
 that $d_c=2$. 
 On the other hand, the higher the dimensionality, the better our
 estimates.  
In the limit of infinite dimension, 
after the usual rescaling (2.3) of the temperature,
 from (4.5) 
one obtains $\b_c = 1$
which is the critical temperature of the SK model.

It is possible to improve the estimate of the critical temperature
 in a systematic way 
by considering larger cluster instead of a cross made of a single central 
spin  $\s_0$ and of its $2d$ neighbors $\s_k$,
 as we shell discuss in the conclusions.

 For instance,  we have considered a plaquette of four spins plus
 the $8 \, (d-1)$ spins that are their nearest neighbors in 
 the $\pm J$ model. 
 Applying our Bethe-Peierls ansatz to the replicated 
plaquettes, the equation for the critical temperature $T_c(d)$ 
 is again   given by the solution of a rational function
 of $t\equiv \tanh(\beta)$.
 After a lengthy but trivial calculation one has
$$
1+(2d-3)t^4-2(d-1)t^2
+2(d-1)(t^4-t^2) \, t^2 \, \left[
{ (1+t^2)^4 +(1+t^4)^2 \over (1+t^2)^2 \, (1+t^4)^2}
\, + \, 
{2t^2\over (1+t^4)^2} 
\right] \, = \, 0 
\eqno (4.6)
$$
In this case, the lower critical dimension is $d_c=15/11$ instead of $d_c=1$ 
found for the cross. The critical temperature $T_c(d)$
  obtained by (4.6) is shown in fig 3, too.

In our opinion, looking at increasingly larger clusters
it is possible to determine the critical temperature of a 
 $d$-dimensional spin glass 
as the zeros of rational functions of $t=\tanh(\beta)$
reaching an accuracy much larger than 
 that one given by direct numerical methods.
 Moreover, one can also hope to find a  converging sequence
 of lower critical dimensions, simply considering the zeros of the
rational functions with $t=1$  (i.e. $T_c=0$).
 
\bigskip
\noindent
{\bf 5. Conclusions and perspectives }
\medskip
\noindent
The properties of finite dimensional Ising spin glasses are 
largely unknown. The lower critical dimension itself is not known 
although most of numerical simulations 
indicate $d=3$ as the lowest dimension which exhibits a glassy phase at
finite temperature. Furthermore, even if the glassy phase is present,
the existence of
replica symmetry breaking at low dimensionality is still controversial.
All that is a clear indication of the difficulties
encountered when one tries to extract information 
directly from the model.

In our approach we simplify the task. 
Indeed, when we assume replica symmetry, 
our approximation reduces itself  to the study of the model
$$
\Phi = \sum_{\{s_{cr}\} }  \prod_{k=1}^{2d} \, 
\exp \left( \b \, J_k \,   
 \s_0 \, \s_k \, + \, \sqrt{\mu}   \, \omega_k   \s_k
\right) 
\eqno(5.1)
$$
where both the $\omega_k$ and the $J_k$ are quenched random variables.
It should be noticed that in this model, one only  
deals with  $2d+1$ spins and $4d$ quenched variables at most.
The model is completed by the self-consistency equation (3.9)
that we rewrite here as
$$
\avr{<\s_0>^2}={ \sum_k \avr{<\s_k>^2} \over \sum_k 1}
\eqno(5.2)
$$
where the second term of (5.2) is the mean of the overlap on
the lateral spin.

The validity of our approach stems from the possibility of a systematical
improvement.
Following a standard technique we can replace (5.1) by 
$$
Z = \sum_{\{s\}}  \,  \prod_{l,l'} \, 
\exp \left( \b \, J_{ll'} \,   \s_l \, \s_{l'} \right) 
\,  \prod_{l,k} \, 
\exp \left( \b \, J_{lk} \,   
 \s_l \, \s_k \right) \,
 \prod_{k} 
\exp \left( \sqrt{\mu} \, \omega_k   \s_k
\right) 
\eqno(5.3)
$$
where the first product is on all the 
first neighbor spins of a hypercube,
the second product is on the couples of spins
formed by lateral spins labelled by $k$ 
on the faces of the hypercube and their first neighbors, 
 the third product is 
simply on the lateral spins.
This model is completed by the self-consistency equation 
$$
{ \sum_l \avr{<\s_l>^2} \over \sum_l 1}
={ \sum_k\avr{<\s_k>^2} \over \sum_k 1}
\eqno(5.4)
$$
where the first sum runs on all the spin of the hypercube
and the second one on the lateral spins.
The linear dimension of the hypercube can be progressively 
increased, and one expects to converge to the right result in the
limit of large hypercubes. 
 In our opinion, this might be a powerful numerical tool to
 determine the internal energy and the critical temperature
 of a spin glass, superior to a direct approach by Montecarlo simulations. 

Let us also mention the major open problem from a theoretical point of view.
It is the search of the solution of the Bethe-Peierls equations
with replica symmetry breaking,
 to see whether  the unphysical behavior of the internal energy
 at low temperature disappears as it happens in the Parisi
 solution [12] of the SK model.  A first step can be reached
 by looking for a solution with only one breaking.
This can pave the way to the comprehension of the glassy transition 
 at finite dimension. 

\bigskip
\noindent
{\bf ACKNOWLEDGEMENTS}
\medskip
\noindent
We acknowledge the financial support 
 ({\it Iniziativa Specifica} FI3) of the I.N.F.N., 
  National Laboratories  of Gran Sasso.
MS thanks the
 Institut Sup\'erieure Polytechnique 
 de Madagascar for hospitality. 
 We are very grateful to G. Parisi,  J. Raboanary, 
 F. Ricci and  A. Vulpiani 
for useful discussions, suggestions and criticisms.
\vfill\eject
\noindent
{\bf References}
\bigskip
\bigskip
\item{ [1]}           
H. Bethe 
Proc. R. Soc. London {\bf 150}, 552 (1935) 
\bigskip     
\item{ [2]}           
R. Peierls
Proc. R. Soc. London {\bf 154}, 207 (1936) 
\bigskip
\item{ [3]}
D. Sherrington and S. Kirkpatrick,
 Phys. Rev. Lett. {\bf 32}, 1792 (1975)
\bigskip
\item{ [4]}
M. Mezard, G. Parisi and M. Virasoro, {\it Spin glass theory and beyond},
 World Scientific Singapore 1988
\bigskip
\item{[5]}
G. Paladin, M. Pasquini, and M. Serva
J. de Phys. I {\bf 5}, 337 (1995);
 I. J. Mod. Phys. B {\bf 9} 399 (1995)
\bigskip     
\item{[6]}
G.  Paladin and M. Serva
 J. Phys. {\bf A29} 1381 (1996);
 M. Serva, G. Paladin and J. Raboanary,
 Physical Review E, {\bf 52},  R9 (1996). 
\bigskip 
\item{ [7]}
L. Buzano,  A. Maritan,  A. Pelizzola
J. Phys. C {\bf  6}, 327 (1994).
\bigskip 
\item{[8]} 
L. Saul and M. Kardar,
Phys. Rev. E {\bf 48}, 48 (1993) 
\bigskip     
\item{[9]} 
L. Viana and J. Bray, 
J. Phys. C {\bf 18}, 3037 (1985) 
\bigskip     
\item{[10]} 
M. Mezard and G. Parisi
Europhys. Lett. {\bf 3}, 10 (1987) 
\bigskip     
\item{[11]} 
R.N. Bhatt and A. P. Young,
Phys. Rev. B {\bf 37}, 5606 (1988) 
\item{} 
D. Badoni, J.C. Ciria, G. Parisi, F. Ritort, J. Pech and 
J.J. Ruiz-Lorenzo, Europhys. Lett {\bf 21}, 495 (1993) 
\item{} 
J. Wang  and A. P. Young, 
J. Phys.  A {\bf 26}, 1063 (1993) 
\item{} 
F. Ricci,  Thesis of the University `La Sapienza' (1995).
\bigskip     
\item{[12]} 
G. Parisi, 
J. Phys.  A {\bf 13}, L-115,  (1980) 
\vfill\eject
\vskip 0.8truecm
\centerline {\bf Figure Captions}
\vskip 0.5truecm
\noindent
\item{Fig. 1}
Replica symmetry solution of the self-consistency equation
 $\mu^*/(2d\b^2)$ as function of the rescaled temperature $T/ \sqrt{2d}$ 
 for the $\pm J$ model at $d=2,3,4,6$.
The larger the dimension, the higher the corresponding
 line.
The dashed line indicates the infinite dimensional limit
 (overlap of the SK model). 
\bigskip
\item{Fig. 2}
Annealed internal energy $U_a/d=\tanh(\beta)$ (dashed line) and the
Bethe-Peierls solutions $U_{BP}/d$ (full lines)  versus temperature
  $T=\beta^{-1}$
 for the $\pm J$ model at 
 $d=2,3,4,6$. The larger the dimension, the higher the corresponding
 line.
The dotted lines are the estimates of the ground state energy
obtained by imposing that 
$T^2 \, \mu^*$ is a non-decreasing function of the temperature.
\bigskip
\item{Fig. 3}
Rescaled critical temperature $T_c/\sqrt{2d}$ 
versus the dimension $d$.
The Bethe-Peierls solution for the $\pm J$ model
 given by (4.5) is indicated by a full line.
 The improved estimate obtained by (4.6)
 where the cluster is a plaquette of four spins 
 instead of a central spin is indicated by a dashed line.
The squares are the numerical values of  $T_c/\sqrt{2d}$ 
 for $d=2,3,4,6$ joined by a dotted line.
 \bye